\documentclass[preprint]{elsarticle}

\usepackage{amssymb}
\usepackage{graphicx}
\usepackage{dcolumn}% Align table columns on decimal point

\begin{document}

\title{The band-gap structures and recovery rules of generalized 
$n$-component Fibonacci piezoelectric superlattices} 

\author{Da Liu and Weiyi Zhang}

\address{Nanjing National Laboratory of Microstructures and
Department of Physics,  Nanjing University, Nanjing 210093, China}

\begin{abstract}
In this communication, the 
band-gap structures of $n$-CF piezoelectric superlattices have
been calculated using the transfer-matrix-method, the
self-similarity behavior and recovery rule have been systematically
analyzed. Consistent with the rigorous mathematical proof by Hu {\it
et al.}[A.~Hu {\it et al.} Phys. Rev. B. {\bf 48}, 829 (1993)], we
find that the $n$-CF sequences with $2 \le n \le 4$ are identified
as quasiperiodic. The imaginary wave numbers are characterized by 
the self-similar spectrum, their major peaks can all be properly 
indexed. In addition, we find that the $n=5$ sequence belongs to 
a critical case which lies at the border between quasiperiodic to 
non-quasiperiodic structures. The frequency range of self-similarity pattern 
approaches to zero and a unique indexing of imaginary wave numbers 
becomes impossible. Our study offers the information on the critical 
$5$-CF superlattice which was not available before. The classification 
of band-gap structures and the scaling laws around fixed points are 
also given.
\end{abstract}
\maketitle

\section{INTRODUCTION}
From a symmetrical point of view, the solids in nature are
classified as crystals, quasicrystals, and noncrystals\cite{1}. The
crystal structure is characterized by a periodic arrangement of
atoms in space, the waves propagating in crystals have a Bloch wave
form. According to the Bloch theorem all states are extended in
space and eigenenergies form continuous band structures. For fully
disordered noncrystal, it is believed that all states are localized
in space and energy spectra are singular continuous. Quasicrystal is
a unique type of structure which lies at the boundary between
translation invariant crystals and random glassy materials\cite{2}.
Though lacking long-range translational symmetry, it possesses a
certain orientational order. To understand the interim situation
between crystals and noncrystals, much research works have been
carried out on quasicrystals, especially since the experimental
discovery of quasicrystal phase in an Al-Mn alloy with icosahedral
symmetry\cite{3}. The simplest structural model describing 
quasicrystals is the Fibonacci lattice model. It has received a
great deal of attention\cite{4,5,6,7,8,9,10,11,12,13,14,15} since it
contains the basic ingredients of quasicrystals and is relatively
easy to deal with. The Fibonacci lattice model is famous for its
Cantor-set spectrum, the self-similar spectrum is identified as
critical eigenstates which is neither extended as in periodic
system, nor localized as in disordered ones\cite{2}.

To explore the spectral evolution from the translational invariant
crystals to fully disordered materials, a structural model which is
tunable between the two extreme cases are highly desirable. A simple
Fibonacci lattice model is certainly a good candidate since its
symmetry lies in between, but it is not good enough since its order
is not adjustable. In this respect, the generalized Fibonacci
lattice model with $n$-component ($n$-CF) offers an excellent choice
and its order can be tuned by the parameter $n$\cite{16}. The $n$-CF
lattice can be obtained by repeated application of the concurrent
substitution rules $TA_1=A_1A_n$, $TA_i=A_{i-1}$($i=2,\dots,n$),
where $A_1,A_2,\dots,A_n$ are the $n$-components and $T$ is the
substitution operator. The $l$th generation sequence of $n$-CF
lattice is defined as $S_l=T^lA_1$ and the generalized $n$-CF
lattice corresponds to $S_\infty$. The first few generation
sequences are given by $S_0=A_1$, $S_1=A_1A_n$, $S_2=A_1A_nA_{n-1}$,
$\dots$, and $S_n=A_1A_nA_{n-1}\dots A_3A_2A_1$. In general,
$S_l=S_{l-1}+S_{l-n}$($l \ge n$). The generalized $n$-CF lattice
recovers the periodic lattice if $n$ is set to $n=1$, it also
naturally reproduces the standard Fibonacci lattice when $n=2$.
What's unique about the generalized $n$-CF lattice is that it can be
tuned smoothly from periodic, to quasiperiodic, and then finally to
non-quasiperiodic lattices as $n$ changes. As shown by Hu {\it et
al.}\cite{16}, the information on the quasiperiodicity of a $n$-CF
lattice is encoded in the substitution matrix
\begin{equation}
T=\left(
\begin{array}{ccccc}
1      & 0      & 0      & \dots  & 1     \\
1      & 0      & 0      & \dots  & 0     \\
0      & 1      & 0      & \dots  & 0     \\
\vdots & \ddots & \ddots & \ddots & \vdots\\
0      & \dots  & 0      & 1      & 0
\end{array}
\right)_{n\times n},
\end{equation}
whether the $n$-CF lattice is quasiperiodic or not is closely
associated with the Pisot number of characteristic polynomial
equation $\lambda^n-\lambda^{n-1}-1=0$. A Pisot number is present if
the characteristic polynomial equation has only one root $1 <
\lambda^{(n)}_0 < 2$ and all its conjugate roots have their norms
less than 1. Following the Bombier-Taylor theorem, the existence of
Pisot number guarantees that the $n$-CF lattice is quasiperiodic and
can be generated by the cut and projection method\cite{17}. 
Hu {\it et al.}\cite{16} showed that
quasiperiodic lattice exists only if $2 \le n \le 5$; $n=1$
corresponds to a periodic lattice while lattices with $n >5$, though
still ordered, is no longer quasiperiodic. In fact, a close
inspection of the root structure suggests that $n=5$ lattice should
be classified as a critical case since there is a pair of conjugate
roots having their norms equal 1, which neither belongs to the
quasiperiodic case where all conjugate roots have their norms less
than 1, nor belongs to a non-quasiperiodic case where some of the conjugate
roots have their norms larger than 1. The Pisot numbers are
$\lambda^{(2)}_0 =1.61803$, $\lambda^{(3)}_0 =1.46557$,
$\lambda^{(4)}_0 =1.38028$, and $\lambda^{(5)}_0 =1.32472$ for
$n=2,3,4,5$, respectively.

Above discussion suggests that $n$-CF lattice covers a wide range of
lattice types from periodic, quasiperiodic, critical, and non-quasiperiodic
lattices, it offers a natural platform to study the spectra
evolution with structural ordering. Though extensive investigations
have been carried out on the electronic\cite{4,5,6,7,8},
vibrational\cite{9,10,11}, and dielectric\cite{12,13,14,15}
properties of quasiperiodic structures as a vehicle to study the
evolution process from ordered periodic structures to disordered
solids, the most previous studies concentrated on the standard
Fibonacci lattices\cite{4,5,6,7,8,9,10,11,12,13,14,15} and the generalized 2-component Fibonacci 
lattices\cite{18,19,20,21,22,23,24}, the generalized $n$-CF lattices are mostly
discussed in the context of structural properties\cite{25}.
Furthermore, the previous studies mostly dealt with the single
degree problems, and studies on the mode-coupling problems with
energy transfer between different degrees of freedom are only a
few\cite{26,27,28}. Thus, we have carried out a comprehensive study
in this communication on the band-gap structures of $n$-CF piezoelectric
superlattices. The transfer-matrix-method is explored so that the
band-gap structures can be most easily visualized through the
imaginary wave number. Our study show that the band-gap structures
can be classified into a series of big cluster in the frequency
spectrum. For each $n$-CF superlattice, the self-similar spectrum
and recovery rule are present within each generation and across
different generations $l$. Consistent with the rigorous mathematical
proof, we find that the $n$-CF superlattices with $2 \le n \le 4$
can be identified as quasiperiodic. The band-gaps are characterized
by the self-similarity behavior and all the gaps can be properly
indexed; The $n=5$ superlattice belongs to a critical case which
lies at the border between quasi-periodic to non-quasiperiodic structures.
The frequency range of self-similar spectrum approaches zero, a
unique indexing of band-gaps becomes impossible. 

\section{MODEL}
In this communication, we use superlattices made of piezoelectric LiNbO$_{3}$ 
compound as our model systems\cite{26}. To distinguish different 
components of $n$-CF superlattices\cite{16}, each component is made of 
a pair of positively and negatively polarized ferroelectric domains of 
varying lengths ($L^i_+$, $L^i_-$). For simplicity and also for setting 
up length scale, the lengths of positively polarized domains of all 
components are taken to be the same $L_+$, while the lengths of 
negatively polarized domains are used to differentiate the different 
components. A typical superlattice for a given $n$-CF sequence and for 
a given generation $l$ is illustrated in Fig.~1. The experimental 
setting and material parameters are described in detail in the 
references.\cite{27,28}

To calculate the band-gap structures of $n$-CF lattices, we use the 
sequence of $l$th generation as a supercell to form a periodic lattice. 
In this case, the lattice constant $a_l=\sum_{i=1}^{n} F^i_l \bar{L}^i$ 
where $F^i_l$ is the number of $i$th component in $l$th generation 
sequence and $\bar{L}^i=\pi L^i/L_+$ is the reduced length of $i$th 
component\cite{27}. The corresponding reciprocal lattice vectors are 
$K_{m_{l}}=(2\pi/a_l)m_{l}$s with $m_{l}$ denoting an integer, the 
reduced wave number $-\pi/a_l<\bar{k}\le+\pi/a_l$. Thus, the essential 
information on the $l$th generation of $n$-CF sequence is included in 
its Fourier components $\theta(m_{l})$. The band-gap opens at the 
reduced frequency $\bar{\omega} \approx K_{m_{l}}$ if 
$\theta(m_{l}) \neq 0$. For the special case where $\bar{L}^i_+ \equiv \pi$
and $\bar{L}^i_-/\pi$ can all be written as rational numbers, 
$\bar{L}^i=\pi[1+b_-^i/a_-^i]$ with the greatest common divisor 
${\rm gcd}(a_-^i, b_-^i)=1$, the reciprocal lattice vectors can also 
be written as rational numbers
\begin{equation}
K_{m_{l}}=\frac{2 m_{l} \times {\rm lcm}(a_{-}^1,a_{-}^2, \cdots, a_{-}^n)}
{{\rm lcm}(a_{-}^1,a_{-}^2, \cdots, a_{-}^n) \cdot
\sum_{i}^{n} F_l^i (1 + b_-^i / a_-^i )}
\end{equation}
with ${\rm lcm}(a_{-}^1, a_{-}^2, \cdots, a_{-}^n)$ denoting the least 
common multiple. It is easy to shown that $\theta(m_l)$ vanishes when 
$m_{l} ={\rm integer} \times {\rm lcm}(a_{-}^1, a_{-}^2, \cdots, a_{-}^n) 
\cdot \sum_{i}^{n} F_l^i (1 + b_-^i / a_-^i)$. In this case, the band-gap 
structures are divided into big clusters, the self-similarity and scaling 
behavior in each cluster can be used to check whether the given 
superlattice is quasiperiodic or not.

\section{NUMERICAL RESULTS AND DISCUSSIONS}
To study the spectral evolution with generation $l$ for $n$-CF
superlattices, usually the furcation pattern of eigenmodes is traced
to analyze the self-similarity property and to search for the fixed
point. When system involves continuum eigenmode
spectra, it is better to track the furcation pattern of band-gap
instead of the eignemodes since the discrete band-gap structures
encode the essential information of $n$-CF sequences. As the band-gap represents the
forbidden frequencies of eigenmodes, the wave propagating in this
frequency range is attenuated at an imaginary wave number
\rm{Im}$\bar{k}$. For polaritons which involve the
coupling between electromagnetic wave and acoustic wave, the
effective dielectric function in the vicinity of a band-gap always
diverges at low band-gap edge while continuously approaches to zero at
high band-gap edge. This suggests that the asymmetrical peaks
symbolizing imaginary wave numbers can be used to visualize the
band-gap structures. In the following, we shall systematically
analyze the band-gap spectra of $n$-CF superlattices as functions of
generation $l$ and component $n$.

Since the $2$-CF sequence corresponds to the standard Fibonacci
lattices which have been extensively studied before\cite{28}, we
concentrate on $3 \leq n \leq 5$ cases below. In choosing the
components of $n$-CF sequence, we follow the rule so that
$\bar{L}^1>\bar{L}^n>\bar{L}^{n-1} \cdots
>\bar{L}^2$\cite{16}. In Fig.~2, the imaginary wave numbers
Im$\bar{k}$ as functions of frequency $\bar{\omega}$ are presented
for the 9th, 12th, and 16th generations of $3$-CF superlattices. The
three components are chosen as $\bar{L}_-^1=\pi$,
$\bar{L}_-^2=1\pi/3$, and $\bar{L}_-^3=2\pi/3$. The overall patterns
of the band-gap spectra look quite similar among different
generations and the positions of major band-gaps are essentially
independent of generations. Within the same big cluster, the spectra
demonstrate a certain symmetry with respect to the center of big
cluster($\bar{\omega}=3$). In addition, the big cluster is further
subdivided into three smaller clusters and each subcluster has its
spectral center. The spectral detail increases with generation $l$
for a given $n$, thus the self-similarity pattern and scaling law
can be analyzed when $l$ is large enough.

To check the quasiperiodicity of $3$-CF
superlattice, we follow Hu {\it et al.}'s procedure\cite{16} and
identify the major peaks of $n$-CF superlattice. According to the
cut and projection method\cite{17}, the reciprocal lattice vectors
are given by
\begin{equation}
\bar{k}[h_1,h_2,\cdots,h_n]=2D^{-1}\sum_{i=1}^n h_i \eta_i,
\end{equation}
where $D=\sum_{i=1}^n \bar{L^i} \eta_i$ is an average lattice
constant and $h_i$s are any integers. $\eta_i$ is the relative
weight of $i$th component and satisfies
$1/\eta_n=\eta_n/\eta_{n-1}=\cdots=\eta_3/\eta_2=\lambda_0^{(n)}$,
and $\lambda_0^{(n)}$ is the Pisot number of substitution matrix of
$n$-CF sequence. For quasiperiodic superlattices of $n$-CF type ($2
\leq n \leq 5$), $D$ converges quickly for large generation $l$. $D$
does not converge for aperiodic superlattices when $n >5$. Since the
band-gap takes place approximately at $\bar{\omega} \approx K_{m_l}$, 
one can properly identify the various major peaks by using
$K_{m_l}=\bar{k}[h_1,h_2,\cdots,h_n]$. In Fig.~2c, the major peaks
of $3$-CF superlattice of large generation are labeled in this way.
To investigate the self-similarity property of $3$-CF superlattices,
the spectra of imaginary wave numbers are presented in Fig.~3
for the 9th and 16th generations. With 
regarding to both the positions and magnitudes of the peaks, there 
is a clear one-to-one correspondence between them 
in the vicinity of the cluster center. The frequency range of the
self-similar patterns is quite broad extending from
$\bar{\omega}=2.29$ to $\bar{\omega}=3.71$. The dynamical property
of $3$-CF superlattices is recovered after 7-generation and the
scaling parameter is $(\lambda_0^{(3)})^7$. 

Since the peak labeling of band-gaps is one of the essential feature
of quasiperiodic superlattices, it is of interest to see how the
situation goes for the $4$-CF and $5$-CF superlattices. In Fig.~4a, 
the spectrum for 12th generation of $4$-CF superlattice is presented,
the four components are $\bar{L}_-^1=4\pi/3$, $\bar{L}_-^2=1\pi/3$,
$\bar{L}_-^3=2\pi/3$, $\bar{L}_-^4=3\pi/3$. While Fig.~4b is the
similar spectrum for the 12th generation of $5$-CF superlattice with
component setting $\bar{L}_-^1=6\pi/3$, $\bar{L}_-^2=1\pi/3$,
$\bar{L}_-^3=2\pi/3$, $\bar{L}_-^4=4\pi/3$, $\bar{L}_-^5=5\pi/3$.
Labeling the band-gap peaks is a very tedious
job for multi-component Fibonacci superlattices since more component
means denser distribution of reciprocal lattice vectors. For the
$4$-CF superlattice, we are able to identify the major band-gap
peaks with easy. As shown in Fig.~4a, all major peaks are uniquely
labeled as it should be for the quasiperiodic lattice. The situation
becomes more complex for $5$-CF superlattice since it lies at the
border between quasiperiodic and non-quasiperiodic structures. We find that
there exists three or more possible labelings for some major peaks,
and their difference in wave number is less than $10^{-15}$, smaller
than the effective digits that double precision number can offer.
This non-uniqueness seems peculiar to $5$-CF superlattices, we
believe that it may be related to its critical nature of the
structure. 

Unlike the $3$-CF superlattices whose band-gap spectra recovers
every 7th generation around the fixed point at the cluster center,
we find that the spectra for $4$-CF superlattices recovers every
15th generation. Fig.~5 compares the imaginary wave number spectra
for both the 12th generation and expanded view of 27th generation
scaled by $(\lambda_0^{(4)})^{15}$. Due to one more component in
composing the superlattice, the frequency range of self-similar
patterns is significantly reduced in comparison with that of $3$-CF
superlattices. But a clear one-to-one correspondence is still
evident in the narrow range $2.865 < \bar{\omega} < 3.135$ both for peak
positions and magnitudes. If we summarize the recovery rule get so
far for the fixed point near the cluster center, following deductive
formula seems reasonable $\Delta l=2^n-1$. It predicts that the
spectra recover every $\Delta l=1,3,7,15$ for the periodic, standard
Fibonacci\cite{28}, $3$-CF, and $4$-CF superlattices. Of course, 
the above statement holds only for the fixed point near the spectral 
cluster center. As a matter of fact, there are many fixed
points for a given dynamical system\cite{29,30}, and they do not
share the same scaling parameter at all.

If the deductive formula is right, one would expect to find the
band-gap spectra of $5$-CF superlattices recover every 31st
generation, but we did not succeed. Though for 
different settings of components we are able to find an approximate 
fixed point around $\bar{\omega}=3$ for the 3rd, 7th, and 21st 
generation recovery rule, none of them is perfect in the sense that 
either the fixed point is not strictly followed or the frequency 
range of self-similar spectra around the fixed point is too narrow. 
Thus, we are led to 
the conclusion that the critical $5$-CF superlattices are characterized 
by two important features: (1) the major peaks cannot be uniquely labeled; 
(2) the frequency range of self-similarity pattern shrinks to zero.

\section{CONCLUSION}
In this communication, the peak labeling and self-similarity
patterns of imaginary wave number spectra are used as essential
tools to cross check the nature of $n$-CF superlattices. For $n=3$
and $n=4$-CF superlattices, major peaks can always be labeled
uniquely. There exists finite frequency range of self-similarity
pattern around the fixed point and the frequency range decreases as
$n$ increases; For the critical case of $5$-CF superlattices, the
unique labeling of major peaks becomes impossible and the frequency
range of self-similarity pattern approaches to zero. Our study 
substantiates the rigorous mathematical proof by Hu {\it et al.}\cite{16}, 
in particular, valuable information is supplemented on the critical 
$5$-CF structures.

\section{Acknowledgments}
This work was supported in part by the National Basic Research 
Program of China (Grant Nos. 2007CB925104, 2010CB923404). We wish 
to acknowledge the partial financial support from the NNSFC under 
Grant Nos.~10774066 and 11021403, and ``Excellent Youth
Foundation''[10025419].

\begin{figure}[p]
\caption{The components and piezoelectric superlattice of $n$-CF
sequence. (a) All components have an identically positively
polarized domain($\bar{L}_+ \equiv \pi$) plus a negatively polarized
one with varying length. (b) An example of $n$-CF superlattice.}
\end{figure}

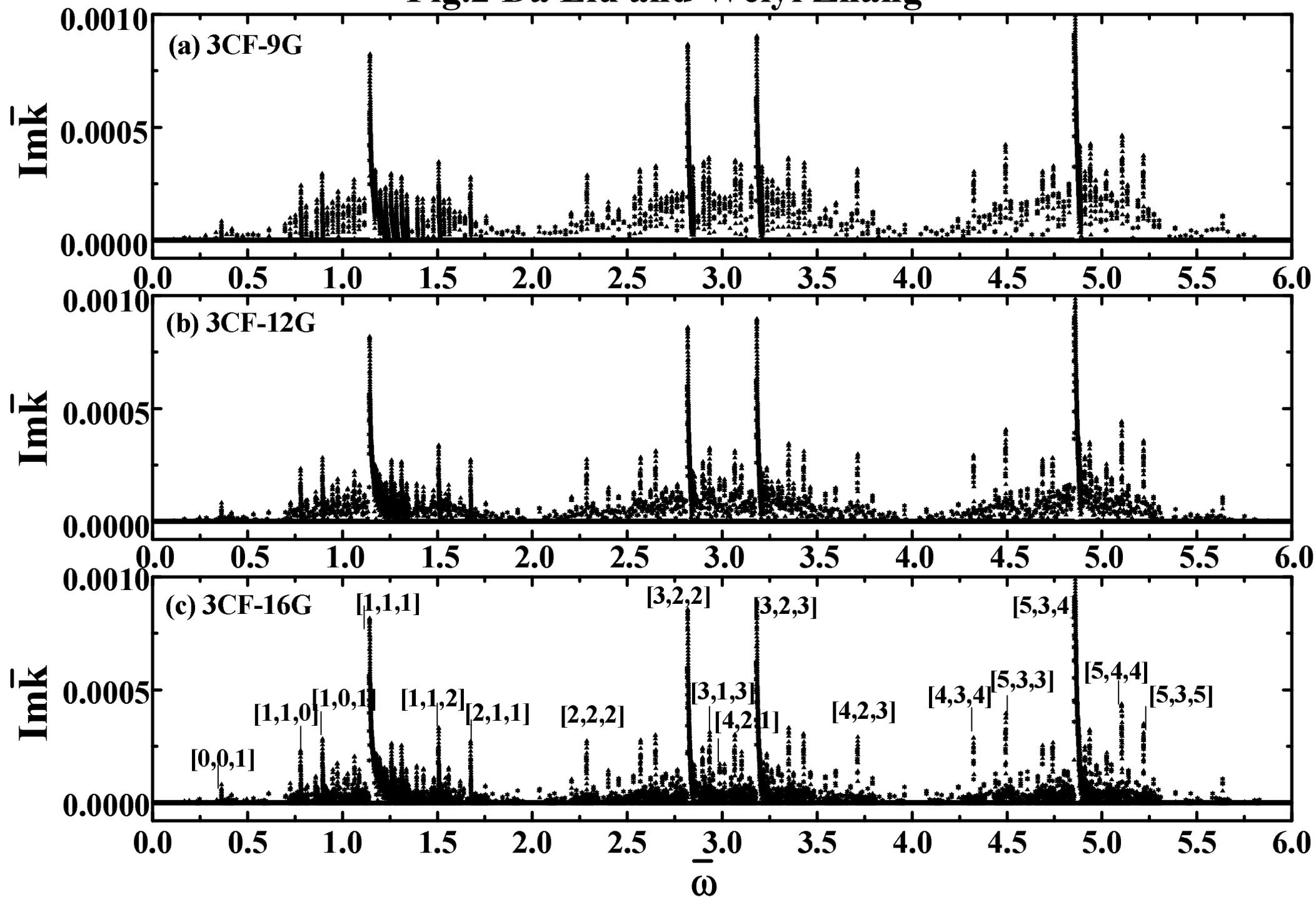
\begin{figure}[p]
\caption{The imaginary wave numbers as functions of frequency for
$3$-CF superlattices. The three components are $\bar{L}_-^1=\pi$,
$\bar{L}_-^2=1\pi/3$, and $\bar{L}_-^3=2\pi/3$. (a) For 9th
generation; (b) For 12th generation; and (c) For 16th generation.
For superlattice of large generation, major peaks can be labeled
according to the projection and cut method.}
\end{figure}

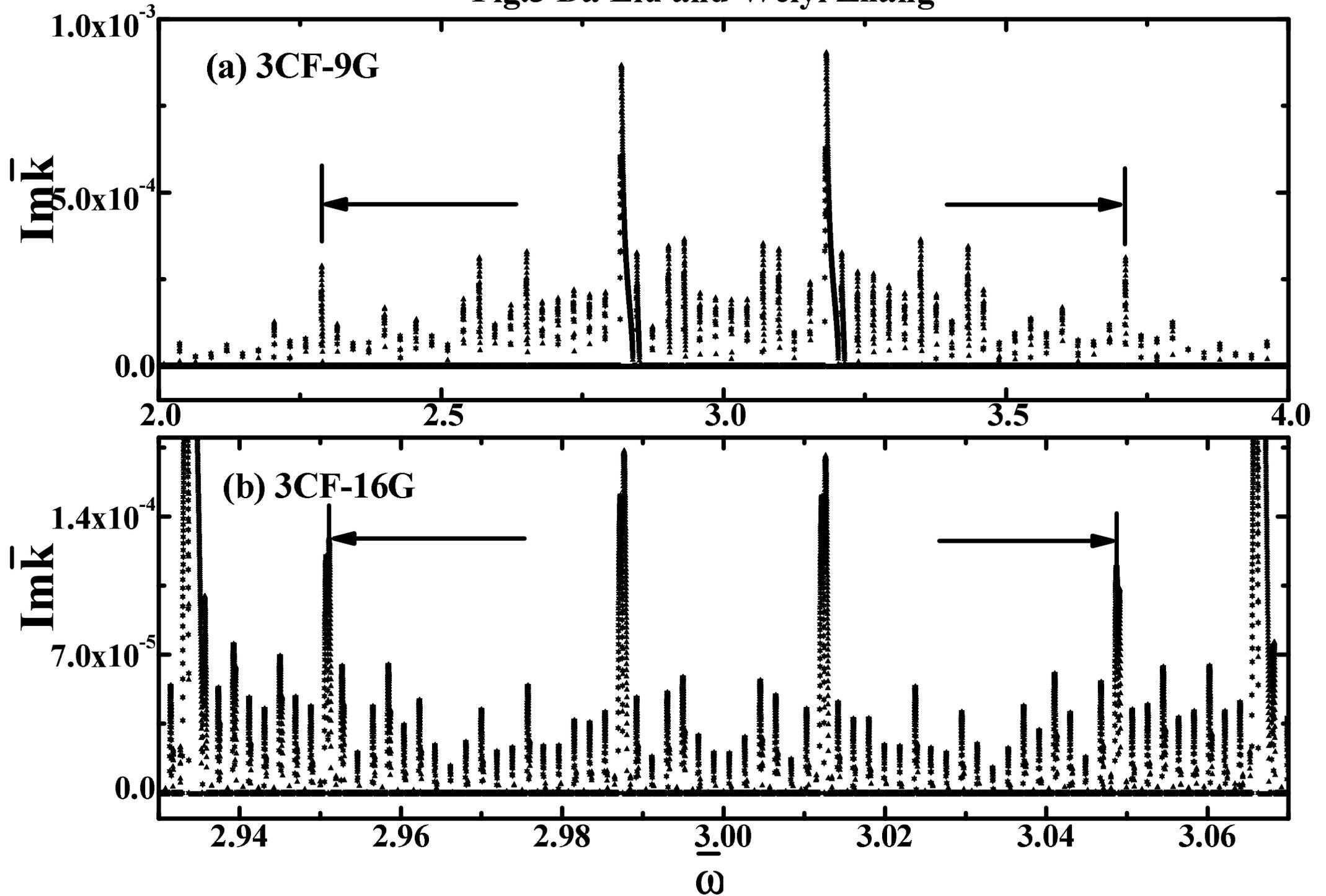
\begin{figure}[p]
\caption{The self-similarity behavior around the fixed point of the
cluster center in $3$-CF superlattices. (a) For 9th generation; (b)
For 16th generation, enlarged around $\bar{\omega}=3$ and scaled by
a factor $(\lambda_0^{(3)})^7$. The other parameter settings are the
same as Fig.~2.}
\end{figure}

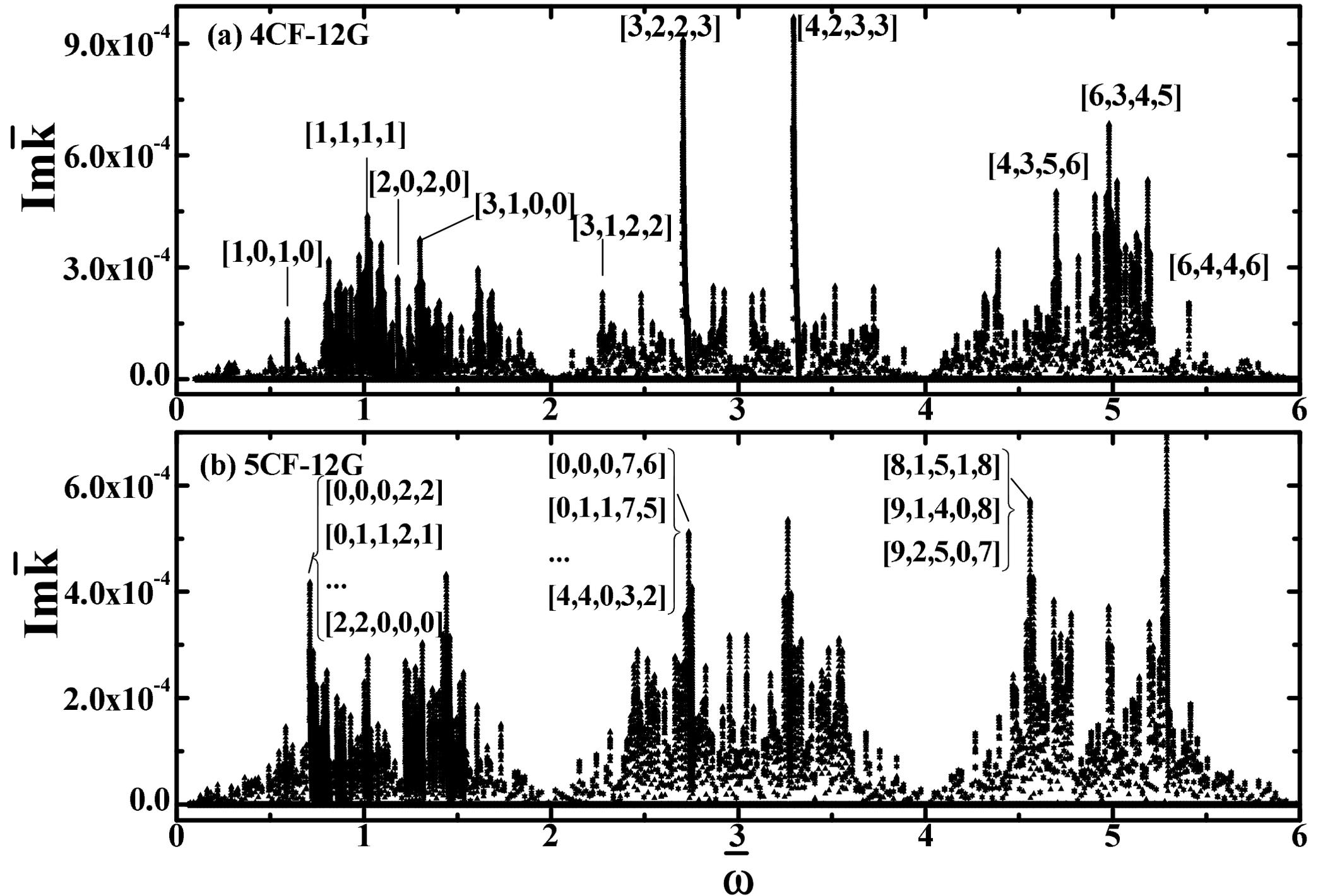
\begin{figure}[p]
\caption{The indexing of major peaks of imaginary wave numbers. (a)
For $4$-CF. The four components are: $\bar{L}_-^1=4\pi/3$,
$\bar{L}_-^2=1\pi/3$, $\bar{L}_-^3=2\pi/3$, $\bar{L}_-^4=3\pi/3$;
(b) For $5$-CF. The five components are: $\bar{L}_-^1=6\pi/3$,
$\bar{L}_-^2=1\pi/3$, $\bar{L}_-^3=2\pi/3$, $\bar{L}_-^4=4\pi/3$,
$\bar{L}_-^5=5\pi/3$. For the critical case $n=5$, the peak labeling
is not unique anymore.}
\end{figure}

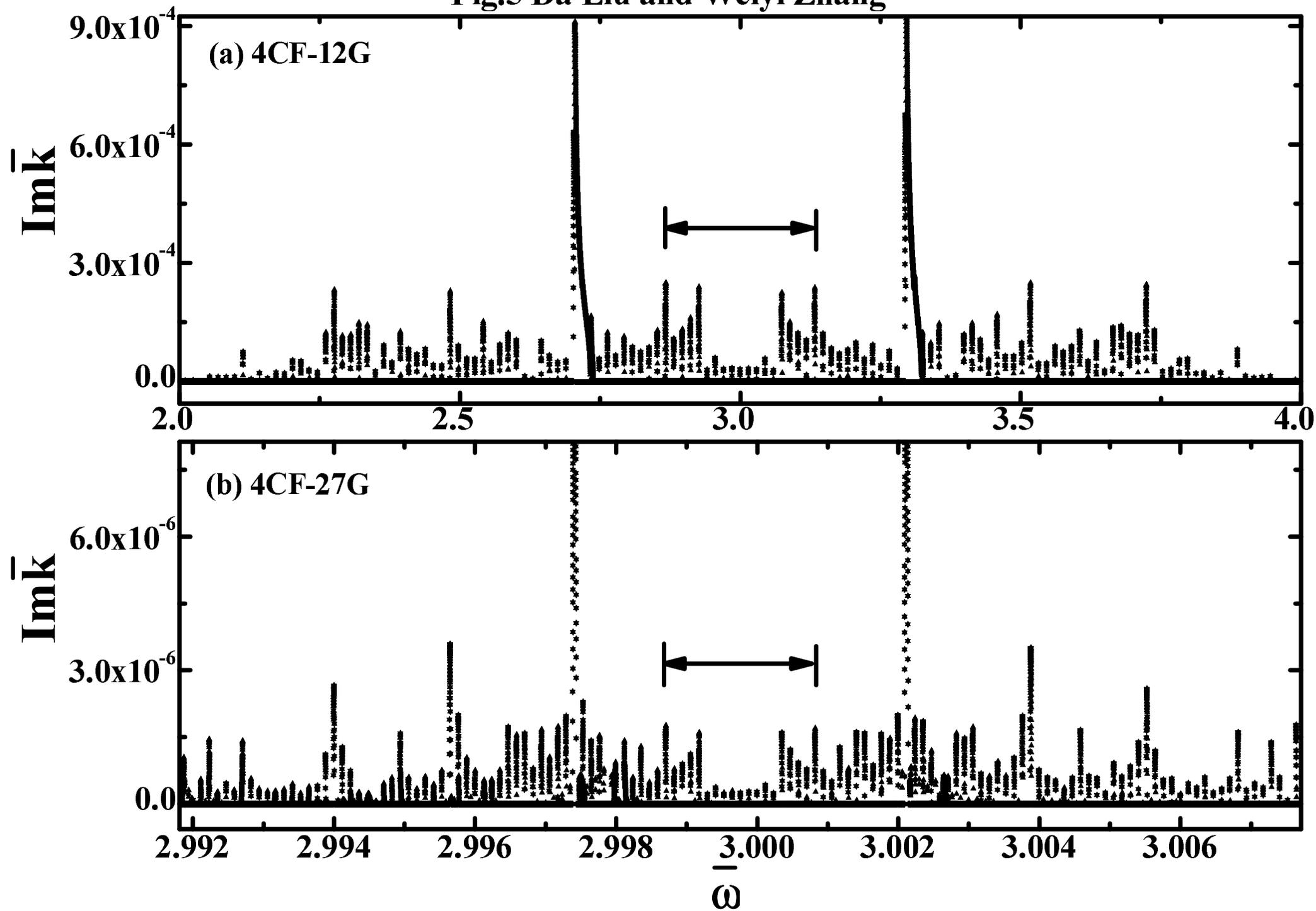
\begin{figure}[p]
\caption{The self-similarity behavior around the fixed point of the
cluster center in $4$-CF superlattices. (a) For 12th generation;
(b) For 27th generation, enlarged around $\bar{\omega}=3$ and scaled
by a factor $(\lambda_0^{(4)})^{15}$. The other parameter settings
are the same as Fig.~4a.}
\end{figure}

\end{document}